# From Technology-Driven Society to Socially Oriented Technology

## The Future of Information Society - Alternatives to Surveillance

by Dirk Helbing (ETH Zurich)

**Our society is changing. Almost nothing these days works without a computer chip. Computing power doubles every 18 months, and in ten years it will probably exceed the capabilities of a human brain. Computers perform approximately 70 percent of all financial transactions today and IBM's Watson now seems to give better customer advise than some human telephone hotlines. What does this imply for our future society?**

The forthcoming economic and social transformation might be more fundamental than the one resulting from the invention of the steam engine. Meanwhile, the storage capacity of data grows even faster than the computational capacity. Within a few years, we will generate more data than in the entire history of humankind. The "Internet of Things" will soon network trillions of sensors together - fridges, coffee machines, electric toothbrushes and even our clothes. Vast amounts of data will be collected. Already, Big Data is being heralded as the oil of the 21st Century.

But this situation will also make us vulnerable. Exploding cyber-crime, economic crises and social protests show that our hyper-connected world is destabilizing. However, is a Surveillance Society the right answer? When all our Internet queries are stored, when our purchases and social contacts are evaluated, when our emails and files are scanned for search terms, and when countless innocent citizens are classified as potential future terrorists, we must ask: Where will this lead to? And where will it end?

Will surveillance lead to self-censorship and discrimination against intellectuals and minorities, even though innovation and creative thinkers are bitterly needed for our economy and society to do well in our changing world? Will free human expression eventually be curtailed by data mining machines analyzing our digital trails?

What are the consequences, say if even the Swiss banks and the U.S. government can no longer protect their secrets, or if our health and other sensitive data is sold on? Or if politically and commercially sensitive strategies can be monitored in real time? What if insider knowledge can be used to undermine fair competition and justice?

The recent allegations that information agencies of various states snoop secretly into the activities of millions of ordinary people has alarmed citizens and companies alike. The moral outrage in response to the surveillance activity has made it clear that it is not a technology-driven society that we need, but instead, a socially-oriented technology, as outlined below. We must recognize that technology without consideration of ethical issues, or without transparency and public discussions can lead us astray. Therefore a new approach to personal data and its uses is required so that we can safely benefit from the many new economic and social opportunities that it can provide.

First, we need a public ethical debate on the concepts of privacy and ownership of data, even more urgently than in bioethics. Important questions that we have to ask are: How do we create opportunities arising in the information age for all, but yet still manage the downside risks and challenges - from cyber-crime to the erosion of trust and democratic rights? Do we really need so much security that we must be afraid of data mining algorithms flagging the activities of millions of ordinary people as suspicious? And what kinds of new institutions would we need in the 21 century?

In the past we have built public roads, parks and museums, schools, libraries and universities. Now, more than ever, we need strategies that protect us against the misuse of data, and that are intended to create transparency and trust. These strategies must place citizen benefits and rights of self-determination at the very core. In addition, we must develop new institutions to provide oversight and control of the new challenges brought by the data revolution. Here are some concrete institutional proposals:

**Self-determined use of personal data:** Already some time ago, the World Economic Forum (WEF) called for a "New Deal on Data" (http://www.weforum.org/pdf/gitr/2009/gitr09fullreport.pdf). It stated that the sustainable use of the economic opportunities of personal data requires a fair balance between economic, governmental and individual interests. A solution would be to return control over personal data to the respective individuals, i.e. give people ownership of their data: the right to possess, access, use and dispose. In addition, individuals should be able to participate in their economic profits. This would require new data protocols and the support of legislation.

**Trusted information exchange:** As the vulnerability of existing systems and the proliferation of cyber-crime indicates, a new network architecture is urgently needed. The handling of sensitive data requires secure encryption, anonymisation and protected pseudonyms, decentralized storage, open software codes and transparency on the use of data, correction possibilities, mechanisms of forgetting, and a protective "digital immune system."

**Credibility mechanisms:** Social mechanisms such as reputation, as seen in the evaluation of information and information sources on the internet, can play a central role in reducing abuse. But remember that the wisdom of crowds only works if individual decisions are not manipulated. Therefore, to be effective, individuals must be given control over the recommendation mechanisms, data filtering and search routines they use, such that they can take decisions based on their own values and quality criteria.

**Participatory platforms:** All over the world people desire increased participation, from consumption to production processes. Now, modern technology allows for the direct social, economic, and political participation of engaged individuals. A basic democracy approach as in Switzerland, where people can decide themselves about many laws, not just political representatives, would be feasible on much larger scales. We also witness an economic trend towards local production, ranging from solar panels to 3D Printers. It can be become a good complement of mass production.

**Open Data:** The innovation ecosystem needs open data and open standards to flourish. Open data enable the rapid creation of new products, which stimulates further products and services. Information is the best catalyst for innovation. Of course, data providers should be adequately compensated, and not all data would have to be open.

**Innovation Accelerator:** To keep pace with our changing world, we need to reinvent the innovation process itself. A participatory innovation process would allow ideas to be implemented faster and external expertise to be integrated more readily. Information is an extraordinary resource: it does not diminish when shared, and it can be infinitely reproduced. **Why shouldn't we use this opportunity?**

**Social Capital:** Information systems can support diverse types of social capital such as trust, reputation, and cooperation. Based on social network interactions, they are the foundation of a flourishing economy and society. So, let's create new value!

**Social Technologies:** Finally, we must learn to build information systems that are compatible with our individual, social and cultural values. We need to design systems that respect the privacy of citizens and prevent fear and discrimination, while promoting tolerance, trust, and fairness. What solutions can we offer users to ensure that information systems are not misused for unjustified monitoring and manipulation? For a well-functioning society, socio-diversity (pluralism) must be protected as much as biodiversity. Both determine the potential for innovation.

These are just some examples of the promising ways in which we could use the Internet of the future. Among all these, a surveillance society is probably the worst of all uses of information technology. A safe and sustainable information society has to be built on reputation, transparency and trust, not mass surveillance.

If we can no longer trust our phones, computers or the Internet, we will either switch off our equipment or start to behave like agents of a secret service: revealing as little information as possible, encrypting data, creating multiple identities, laying false traces.

Such behaviour would create little benefits for ordinary citizens, besides protection, but might help criminals to hide. It would be a pity if we failed to use the opportunities afforded by the information age, just because we did not think hard or far enough about the technological and legal frameworks and institutions needed.

The information age is now at a crossroad. It may eventually lead us to a totalitarian surveillance state, or we can use it to enable a creative, participatory society. It is our decision, and we should not leave it to others.

It is also time to build the institutions for the globalized information society to come, in a world-wide collaboration, instead of starting a global war of information systems.

## Further Reading